# A Game-Theoretic Approach for Runtime Capacity Allocation in MapReduce

Eugenio Gianniti[*], Danilo Ardagna[†], and Michele Ciavotta[‡]

Dipartimento di Elettronica, Informazione e Bioingegneria,
Politecnico di Milano
Milano, Italy

Mauro Passacantando[§]

Dipartimento di Informatica,
Università di Pisa
Pisa, Italy

**Abstract**

Nowadays many companies have available large amounts of raw, unstructured data. Among Big Data enabling technologies, a central place is held by the MapReduce framework and, in particular, by its open source implementation, *Apache Hadoop*. For cost effectiveness considerations, a common approach entails sharing server clusters among multiple users. The underlying infrastructure should provide every user with a fair share of computational resources, ensuring that Service Level Agreements (SLAs) are met and avoiding wastes.

In this paper we consider two mathematical programming problems that model the optimal allocation of computational resources in a Hadoop 2.x cluster with the aim to develop new capacity allocation techniques that guarantee better performance in shared data centers. Our goal is to get a substantial reduction of power consumption while respecting the deadlines stated in the SLAs and avoiding penalties associated with job rejections. The core of this approach is a distributed algorithm for runtime capacity allocation, based on Game Theory models and techniques, that mimics the MapReduce dynamics by means of interacting players, namely the central Resource Manager and Class Managers.

---

[*]`eugenio.gianniti@polimi.it`
[†]`danilo.ardagna@polimi.it`
[‡]`michele.ciavotta@polimi.it`
[§]`mauro.passacantando@unipi.it`

# 1 Introduction

A large number of enterprises currently commits to the extraction of information from huge data sets as part of their core business activities. Applications range from fraud detection to one-to-one marketing, encompassing business analytics and support to decision making in both private and public sectors. In order to cope with the unprecedented amount of data and the need to process them in a timely fashion, new technologies are increasingly adopted in industry, following the *Big Data* paradigm. Among such technologies, Apache Hadoop [1] is already widespread and predictions suggest a further increase in its future adoption. IDC estimates that, by 2020, nearly 40% of Big Data analyses will be supported by public Clouds [2], while Hadoop touched half of the data worldwide by 2015 [3].

Apache Hadoop is an open source software suite that enables the elaboration of vast amounts of data on clusters of commodity hardware. Hadoop implements the MapReduce paradigm, automatically ensuring parallelization, distribution, fault-tolerance, reliability, and monitoring. In order to obtain a high level of scalability, Hadoop 2.x overcomes the drawbacks present in the previous versions implementing a distributed resource management system, with a central Resource Manager (RM) that provides resources for computation to Application Masters (AMs) entitled to manage the submitted jobs.

Despite the convenience of this paradigm and the undeniably widespread adoption of Hadoop within the IT industry, still there are no tools that support developers and operators in achieving optimal capacity planning of MapReduce applications. In this context the main drawback [4], [5] is that the execution time of a MapReduce job is generally unknown in advance: for some systems, capacity allocation can become a critical aspect. Moreover, resource allocation policies need to decide job execution and rejection rates in a way that users' workloads can meet their Service Level Agreements (SLAs) and the overall cost is minimized.

This paper investigates the theoretical foundations for the optimal runtime management of cluster resources in private Clouds. We envisage a scenario where a novel resource allocation policy, based on our findings, is implemented and adopted in order to optimally address the discussed issues. Precisely, we focus on the joint admission control and capacity allocation problem, seeking to fulfill SLAs while minimizing energy-related costs. Overall, ICT energy demand sums up to 7% of the world consumption and was expected to rise up to 12% by 2017 [6], with a further tendency towards a shift from devices to networks and data centers consumption [7]. Indeed, worldwide ICT systems account for 2–4% of global $CO_2$ emissions and it is expected that they can reach up to 10% in 5–10 years [8].

We propose a theoretical approach in which the allocation problem is solved periodically based on a prediction of the forthcoming system load. In particular, we adopt Game Theory techniques, which found successful application in the field of Cloud computing [9]–[12], and use them to provide a distributed, scalable solution to the joint admission control and capacity allocation of multi-class Hadoop clusters. We propose a distributed solution leading to a Generalized Nash Equilibrium Problem (GNEP), a class of games that generalizes classical Nash problems, yielding much more difficult instances.

This paper is organized as follows. Initially, we give a clear statement of the problem at hand alongside relevant design assumptions, in Section 2. Afterwards, we show how we developed models to solve the joint capacity allocation and admission control problem. Section 3 presents a preliminary, centralized mathematical programming formulation, whilst Section 4 builds on it to propose a distributed game-theoretic model. Then we analyze our results in Section 5, whilst Section 6 discusses other literature proposals. In the end, Section 7 wraps up this work and draws conclusions on the outcomes.

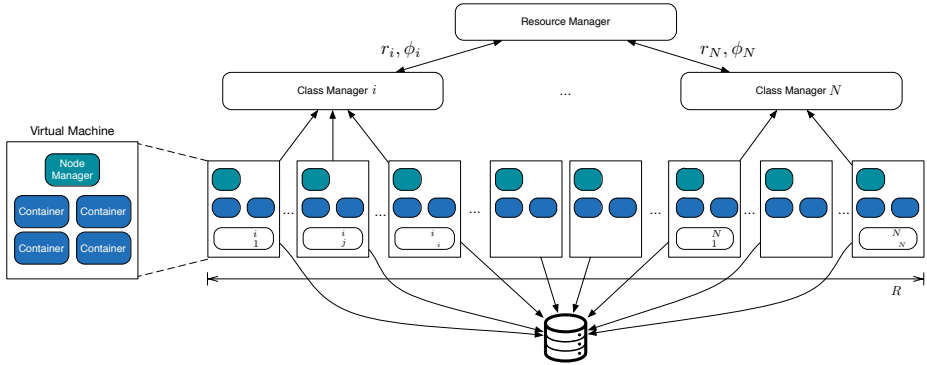

Figure 1: Reference technology

## 2 Problem Statement and Design Assumptions

Figure 1 shows the reference technological system, featuring the Hadoop 2.x framework running on a private Cloud. The private Cloud cluster supports several user classes competing for resources, which are managed via the YARN Capacity Scheduler. Each class collects similar jobs, i.e., applications that share analogous values for parameters characterizing their performance: they have the same *job profile*. Following the notation brought forth in [5], [13], job profiles include the following contributes: $n_i^M$ and $n_i^R$, the total number of Map and Reduce tasks per job, respectively; $M_i^{max}$, $R_i^{max}$, $Sh_{1,i}^{max}$, and $Sh_{typ,i}^{max}$, the maximum durations of one single Map, Reduce, and Shuffle task (notice that the first Shuffle wave of a given job is distinguished from all the subsequent ones); $M_i^{avg}$, $R_i^{avg}$, and $Sh_{typ,i}^{avg}$, i.e., the average duration of Map, Reduce, and Shuffle tasks, respectively.

The modeled cluster supports the concurrent execution of a maximum of $R$ virtual machines (VMs), which we assume homogeneous for the sake of simplicity. In order to allow for elasticity, the reference system does not store data on the Hadoop Distributed File System (HDFS) as this would expose it to data corruption or poor performance. On the contrary, according to the practice suggested by major Cloud providers [14], [15], data reside on external storage [16], [17].

According to our vision of a novel resource allocation policy, every application class is managed by a Class Manager (CM), which negotiates the required resources with a central RM, entitled to split the available capacity among submitted jobs. The set of application classes is denoted with $\mathcal{A}$ and $N = |\mathcal{A}|$. For all CMs $i \in \mathcal{A}$, the RM assigns $r_i$ VMs. In other words, in this scenario the proposed framework acts as the YARN Capacity Scheduler [18], assigning every application class $i$ to a separate queue and providing a portion $\phi_i$ of the total resources, where:

$$\phi_i \triangleq \frac{r_i}{\sum_{j=1}^{N} r_j}, \ \forall i \in \mathcal{A}.$$

Given $\bar{\rho}$, the time unit cost to run a single VM, it is possible to obtain the total cost of execution as $\sum_{i=1}^{N} \bar{\rho} r_i$.

For every application class $i$, an SLA establishes that a maximum of $H_i^{up}$ jobs can be executed concurrently. However, the system can autonomously decide to reject a portion of such jobs upon payment of a penalty. Finally, the accepted $h_i$ jobs cannot be fewer than $H_i^{low}$ and the system commits to complete them within a deadline $D_i$. We denote with $\mathcal{P}_i(h_i)$ the penalty functions associated to the possible rejection of some jobs. They are assumed to be decreasing and convex: this is reasonable as it means that penalties increase at least linearly in the number of rejected jobs.

According to the obtained number of resources $r_i$, a CM may need to reject some jobs, then it proceeds to activate a suitable number of AMs to coordinate the admitted ones. In this scenario, the AMs have the only duty of managing the resources obtained by the CMs so as to carry out the associated job tasks, without directly taking part in the allocation process.

We propose to solve our problems hourly, based on a prediction of the load $H_i^{up}$, to dynamically reallocate resources among application classes, while also avoiding the overhead and costs of booting and shutting down VMs too frequently.

In the Hadoop framework each computational node hosts several slots that execute Map and Reduce tasks. In particular, according to the YARN configuration, VM resources are split in containers, so that every VM can be used to concurrently run $c_i^M$ Map or $c_i^R$ Reduce tasks[1]. These parameters depend only on the job classes owing to the assumption of homogeneity made on VMs. The total number of Map and Reduce slots assigned to an application class is represented by $s_i^M$ and $s_i^R$, respectively. Again, these variables give a simple representation of the workload required to complete jobs in each class due to the homogeneity assumption on VMs. Precisely, with $s_i^M$ and $s_i^R$ we represent the number of Map and Reduce tasks that run concurrently, hence the maximum size of each wave.

As previously stated, according to [5], it is possible to derive from the Hadoop logs a *job profile*, i.e., a set of parameters that characterize the execution of jobs in each class. In this paper we use a more refined formulation, as in [13]. The estimated minimum and maximum execution times are computed with formulae similar to the following:

---

[1] Note that in Hadoop 1.x, each node resources can be partitioned between slots assigned to Map tasks and slots assigned to Reduce tasks. In Hadoop 2.x, the resource capacity configured for each container is available for both Map and Reduce tasks and cannot be partitioned anymore [19]. The maximum number of concurrent mappers and reducers (the *slot count*) is calculated by YARN based on administrator settings [20]. A node is eligible to run a task when its available memory and CPU can satisfy the task resource requirement. With our hypothesis above, we assume that the configuration settings are such that whatever combination of Map and Reduce tasks can be executed within a container, no CPU remains idle because of a wrong setting of these parameters.

$$T_i = A_i \frac{h_i}{s_i^M} + B_i \frac{h_i}{s_i^R} + C_i. \tag{1}$$

The parameters $A_i$, $B_i$, and $C_i$ aggregate the already mentioned $n_i^M$, $n_i^R$, $M_i^{max}$, $R_i^{max}$, $Sh_{1,i}^{max}$, $Sh_{typ,i}^{max}$, $M_i^{avg}$, $R_i^{avg}$, and $Sh_{typ,i}^{avg}$ parameters, which are measured directly by Hadoop and easily obtainable from its execution logs. These formulae are used to predict the jobs execution time, given the number of allocated resources and the concurrency level.

Equations (1) can be used to derive deadline constraints; two main alternatives have to be considered, though. On one hand it is possible to express constraints giving strong guarantees of meeting hard deadlines considering a conservative upper bound estimate. When dealing with soft deadlines, instead, the arithmetic mean of the upper and lower bounds has been shown to be a better suited estimate (see [5], [13]), giving a quite accurate forecast of the actual execution time with just an average 10% gap between predicted and measured times [13], and leading to the allocation of comparably fewer resources. Notwithstanding, in both cases we can formulate the deadline constraints as:

$$T_i = A_i \frac{h_i}{s_i^M} + B_i \frac{h_i}{s_i^R} + C_i \leq D_i, \quad \forall i \in \mathcal{A}. \tag{2}$$

where $D_i$ are the deadlines. In the following, we adopt the parameter $E_i = C_i - D_i$. Notice that, by definition, it holds $E_i < 0$, as nonnegative values would mean that jobs of class $i$ cannot be completed on time. In this paper, we adopt the average formulation, hence renouncing to guarantees that the admitted jobs are completed on time, in favor of a less demanding allocation.

In light of the above, we can say that the ultimate goal of the proposed approach is to determine the optimal values of $h_i$, $s_i^M$, $s_i^R$, and $r_i$ so that the sum of costs and rejection penalties is minimized, while the deadlines set by SLAs are met. In Table 1 are reported all the parameters used in the models discussed in the subsequent sections, while in Table 2 we summarize the decision variables.

## 3 Mathematical Programming Formulation

Building upon the observations and assumptions previously discussed, we formulate a preliminary mathematical programming model that formalizes the problem. The model is the following:

$$\min_{\mathbf{r},\mathbf{h},\mathbf{s}^M,\mathbf{s}^R} \quad \sum_{i=1}^{N} \bar{\rho} r_i + \sum_{i=1}^{N} \mathcal{P}_i(h_i) \tag{P1a}$$

subject to:

$$\sum_{i=1}^{N} r_i \leq R, \tag{P1b}$$

$$H_i^{low} \leq h_i \leq H_i^{up}, \quad \forall i \in \mathcal{A}, \tag{P1c}$$

Table 1: Centralized Model Parameters

| | |
|---|---|
| $\mathcal{A}$ | Set of job classes |
| $N$ | Number of CMs, or $|\mathcal{A}|$ |
| $\bar{\rho}$ | Time unit cost for running a VM in the cluster |
| $H_i^{up}$ | Maximum concurrency required in the SLA contract for job class $i$ |
| $H_i^{low}$ | Minimum concurrency required in the SLA contract for job class $i$ |
| $\psi_i^{low}$ | Reciprocal of $H_i^{up}$ |
| $\psi_i^{up}$ | Reciprocal of $H_i^{low}$ |
| $R$ | Total capacity of the cluster as number of VMs |
| $A_i$ | Coefficient associated to Map tasks in the job profile for job class $i$, [13] |
| $B_i$ | Coefficient associated to Reduce tasks in the job profile for job class $i$, [13] |
| $E_i$ | Parameter lumping the constant terms associated neither to Map, nor to Reduce tasks in the job profile for job class $i$, as well as the deadlines, [13] |
| $c_i^M$ | Map slots supported on one VM for job class $i$ |
| $c_i^R$ | Reduce slots supported on one VM for job class $i$ |
| $\alpha_i$ | Slope of the penalty contribution linear in $\psi_i$ for job class $i$ |
| $\beta_i$ | Constant term of the penalty contribution linear in $\psi_i$ for job class $i$ |

Table 2: Centralized Model Decision Variables

| | |
|---|---|
| $r_i$ | Number of VMs assigned for the execution of job class $i$ |
| $h_i$ | Number of jobs concurrently executed in job class $i$ |
| $\psi_i$ | Reciprocal of the concurrency degree $h_i$ |
| $s_i^M$ | Number of Map slots assigned for the execution of job class $i$ |
| $s_i^R$ | Number of Reduce slots assigned for the execution of job class $i$ |

$$\frac{A_i h_i}{s_i^M} + \frac{B_i h_i}{s_i^R} + E_i \leq 0, \quad \forall i \in \mathcal{A}, \tag{P1d}$$

$$\frac{s_i^M}{c_i^M} + \frac{s_i^R}{c_i^R} \leq r_i, \quad \forall i \in \mathcal{A}, \tag{P1e}$$

$$r_i \in \mathbb{N}, \quad \forall i \in \mathcal{A}, \tag{P1f}$$

$$h_i \in \mathbb{N}, \quad \forall i \in \mathcal{A}, \tag{P1g}$$

$$s_i^M \in \mathbb{N}, \quad \forall i \in \mathcal{A}, \tag{P1h}$$

$$s_i^R \in \mathbb{N}, \quad \forall i \in \mathcal{A}. \tag{P1i}$$

In problem (P1) the objective function (P1a) has a term representing the cost of executing all the assigned VMs and another for penalties. Constraint (P1b) ensures that the cluster capacity bounds the total assigned resources. Further, the set of constraints (P1c) imposes the minimum and maximum job concurrency levels, according to the SLAs. Similarly, constraints (P1d) exploit the job profiles to ensure the deadlines are met. Constraints (P1e) guarantee that every application class receives enough VMs to support the number of slots they should run concurrently. The left hand side is a conservative estimate of the resources needed to support at the same time $s_i^M$ and $s_i^R$ slots: this expression greatly simplifies the analysis. Constraints (P1f)–(P1i) require all the variables to be nonnegative integers, as expected for their interpretation. In particular, notice that the other constraints impose that all the variables must be positive integers.

Since the optimization problem is nonlinear due to the constraints family (P1d) and the penalty terms $\mathcal{P}_i(h_i)$, it is advisable to study its continuous relaxation. Instances of practical interest may well have hundreds of application classes, thus making the solution methods for nonlinear integer problems infeasible for supporting admission control and capacity allocation at runtime. Indeed, the model includes $4N$ integer variables and $8N + 1$ constraints. Nonetheless, the solutions to the proposed models have to be integer, as it is only possible to instantiate a discrete number of VMs, then we will discuss a heuristic approach to the issue in Section 4.5.

Moreover, constraints (P1d) are not convex, thus ruling out many important results for nonlinear optimization. We address this issue introducing a new set of variables $\psi_i \triangleq h_i^{-1}$, so that constraints (P1d) are now convex, as shown in Proposition 3.1.

**Proposition 3.1.** *The function*

$$f\left(\psi_i, s_i^M, s_i^R\right) = \frac{A_i}{s_i^M \psi_i} + \frac{B_i}{s_i^R \psi_i} + E_i$$

*is convex.*

*Proof.* We note that it is sufficient to prove that the function $g(x,y) = \frac{1}{xy}$ is convex whenever $x$ and $y$ are positive. The Hessian matrix of $g$ is:

$$\nabla^2 g(x,y) = \begin{bmatrix} \frac{2}{x^3 y} & \frac{1}{x^2 y^2} \\ \frac{1}{x^2 y^2} & \frac{2}{xy^3} \end{bmatrix}.$$

Since both the trace and the determinant are positive, the Hessian matrix of $g$ is positive definite for any positive $x$ and $y$, hence $g$ is convex. □

According to Proposition 3.1, with this change of variables it is possible to write a convex nonlinear problem. Note that we explicitly impose $\psi_i > 0$ with the rewriting of constraints (P1c) whilst the same does not hold for $s_i^M$ and $s_i^R$. Besides the trivial consideration that $s_i^M = 0$ and $s_i^R = 0$ are outside the domain of $f$, we should notice that this mirrors the fact that not assigning slots to a job class is not acceptable in the modeled system. Now, let us assume that the penalties are linear in the new variables $\psi_i$, hence it is possible to write them as $\alpha_i \psi_i - \beta_i$, $\forall i \in \mathcal{A}$. The corresponding penalty term in the objective function (P1a) is then $\mathcal{P}_i(h_i) = \alpha_i h_i^{-1} - \beta_i$, $\forall i \in \mathcal{A}$. This expression is consistent with the assumptions of convexity and monotonicity made on $\mathcal{P}_i(h_i)$. The formulation reads:

$$\min_{\mathbf{r}, \boldsymbol{\psi}, \mathbf{s}^M, \mathbf{s}^R} \sum_{i=1}^N \bar{\rho} r_i + \sum_{i=1}^N (\alpha_i \psi_i - \beta_i) \tag{P2a}$$

subject to:

$$\sum_{i=1}^N r_i \leq R, \tag{P2b}$$

$$\psi_i^{low} \leq \psi_i \leq \psi_i^{up}, \quad \forall i \in \mathcal{A}, \tag{P2c}$$

$$\frac{A_i}{s_i^M \psi_i} + \frac{B_i}{s_i^R \psi_i} + E_i \leq 0, \quad \forall i \in \mathcal{A}, \tag{P2d}$$

$$\frac{s_i^M}{c_i^M} + \frac{s_i^R}{c_i^R} \leq r_i, \quad \forall i \in \mathcal{A}, \tag{P2e}$$

$$r_i \geq 0, \quad \forall i \in \mathcal{A}, \tag{P2f}$$

$$\psi_i \geq 0, \quad \forall i \in \mathcal{A}, \tag{P2g}$$

$$s_i^M \geq 0, \quad \forall i \in \mathcal{A}, \tag{P2h}$$

$$s_i^R \geq 0, \quad \forall i \in \mathcal{A}. \tag{P2i}$$

Following the proposed change of variables, constraints (P1c) become constraints (P2c) where $\psi_i^{low} = (H_i^{up})^{-1}$ and $\psi_i^{up} = (H_i^{low})^{-1}$. Further, as can be seen from constraints (P2f)–(P2i), we take the continuous relaxation of the otherwise mixed integer problem. Thanks to the high values typically attained by $s_i^M$, $s_i^R$, and $r_i$, it is possible to round the real solution without affecting too much the optimal value.

We now proceed with the analysis of this formulation. Problem (P2) is convex and Slater constraint qualification holds: the Karush-Kuhn-Tucker (KKT) conditions are, then, necessary and sufficient for optimality. The associated Lagrangian is:

$$\mathcal{L}(\mathbf{r}, \boldsymbol{\psi}, \mathbf{s}^M, \mathbf{s}^R) = \sum_{i=1}^{N} \bar{\rho} r_i + \sum_{i=1}^{N} (\alpha_i \psi_i - \beta_i) +$$

$$+ a \left( \sum_{i=1}^{N} r_i - R \right) + \sum_{i=1}^{N} b_i \left( \psi_i^{low} - \psi_i \right)$$

$$+ \sum_{i=1}^{N} c_i \left( \psi_i - \psi_i^{up} \right) +$$

$$+ \sum_{i=1}^{N} d_i \left( \frac{A_i}{s_i^M \psi_i} + \frac{B_i}{s_i^R \psi_i} + E_i \right) + \quad (3)$$

$$+ \sum_{i=1}^{N} e_i \left( \frac{s_i^M}{c_i^M} + \frac{s_i^R}{c_i^R} - r_i \right) - \sum_{i=1}^{N} f_i s_i^M +$$

$$- \sum_{i=1}^{N} g_i s_i^R - \sum_{i=1}^{N} k_i r_i - \sum_{i=1}^{N} l_i \psi_i.$$

The associated KKT conditions are:

$$\frac{\partial \mathcal{L}}{\partial r_i} = \bar{\rho} + a - e_i - k_i = 0, \quad \forall i \in \mathcal{A}, \quad (4a)$$

$$\frac{\partial \mathcal{L}}{\partial \psi_i} = \alpha_i - b_i + c_i - \frac{d_i A_i}{s_i^M \psi_i^2} - \frac{d_i B_i}{s_i^R \psi_i^2} = 0, \quad \forall i \in \mathcal{A}, \quad (4b)$$

$$\frac{\partial \mathcal{L}}{\partial s_i^M} = -\frac{d_i A_i}{\psi_i \left( s_i^M \right)^2} + \frac{e_i}{c_i^M} = 0, \quad \forall i \in \mathcal{A}, \quad (4c)$$

$$\frac{\partial \mathcal{L}}{\partial s_i^R} = -\frac{d_i B_i}{\psi_i \left( s_i^R \right)^2} + \frac{e_i}{c_i^R} = 0, \quad \forall i \in \mathcal{A}. \quad (4d)$$

And the complementary slackness conditions:

$$a \left( \sum_{i=1}^{N} r_i - R \right) = 0, \ a \geq 0, \quad (5a)$$

$$b_i \left( \psi_i^{low} - \psi_i \right) = 0, \ b_i \geq 0, \quad \forall i \in \mathcal{A}, \quad (5b)$$

$$c_i \left( \psi_i - \psi_i^{up} \right) = 0, \ c_i \geq 0, \quad \forall i \in \mathcal{A}, \quad (5c)$$

$$d_i \left( \frac{A_i}{s_i^M \psi_i} + \frac{B_i}{s_i^R \psi_i} + E_i \right) = 0, \ d_i \geq 0, \quad \forall i \in \mathcal{A}, \quad (5d)$$

$$e_i \left( \frac{s_i^M}{c_i^M} + \frac{s_i^R}{c_i^R} - r_i \right) = 0, \ e_i \geq 0, \quad \forall i \in \mathcal{A}, \quad (5e)$$

$$f_i s_i^M = 0, \ f_i \geq 0, \quad \forall i \in \mathcal{A}, \quad (5f)$$

$$g_i s_i^R = 0, \ g_i \geq 0, \quad \forall i \in \mathcal{A}, \quad (5g)$$

$$k_i r_i = 0, \ k_i \geq 0, \quad \forall i \in \mathcal{A}, \quad (5h)$$

$$l_i \psi_i = 0, \ l_i \geq 0, \quad \forall i \in \mathcal{A}. \quad (5i)$$

Now, we can easily prove the following propositions.

**Proposition 3.2.** *Constraints* (P2d) *and* (P2e) *are active in every optimal solution.*

*Proof.* Building upon the previous consideration that all the variables must be positive in feasible solutions and owing to (5h), we have $k_i = 0$. From (4a), then:

$$e_i = \bar{\rho} + a \geq \bar{\rho} > 0, \quad \forall i \in \mathcal{A},$$

meaning that every constraint (P2e) is active in optimal solutions.

Now, conditions (4c) yield:

$$d_i = \frac{e_i \psi_i \left(s_i^M\right)^2}{A_i c_i^M}, \quad \forall i \in \mathcal{A},$$

and, since all the parameters and variables are positive, it is proved that $d_i > 0, \forall i \in \mathcal{A}$, hence all the (P2d) are active in every optimal solution as well. $\square$

**Proposition 3.3.** *The optimal values attained by $s_i^M$, $s_i^R$, and $\psi_i$ in problem* (P2) *are:*

$$s_i^M = \xi_i^M r_i, \quad \forall i \in \mathcal{A}, \tag{6a}$$

$$s_i^R = \xi_i^R r_i, \quad \forall i \in \mathcal{A}, \tag{6b}$$

$$\psi_i = K_i r_i^{-1}, \quad \forall i \in \mathcal{A}, \tag{6c}$$

*where:*

$$\xi_i^M \triangleq \frac{c_i^M}{1 + \sqrt{\frac{B_i}{A_i} \frac{c_i^M}{c_i^R}}}, \quad \forall i \in \mathcal{A}, \tag{7a}$$

$$\xi_i^R \triangleq \frac{c_i^R}{1 + \sqrt{\frac{A_i}{B_i} \frac{c_i^R}{c_i^M}}}, \quad \forall i \in \mathcal{A}, \tag{7b}$$

$$K_i \triangleq -\frac{\left(\sqrt{\frac{A_i}{c_i^M}} + \sqrt{\frac{B_i}{c_i^R}}\right)^2}{E_i}, \quad \forall i \in \mathcal{A}. \tag{7c}$$

*Proof.* From (4c) and (4d) we obtain:

$$s_i^M = s_i^R \sqrt{\frac{A_i c_i^M}{B_i c_i^R}}, \quad \forall i \in \mathcal{A}.$$

Substituting in (P2e) we get:

$$s_i^R = \frac{c_i^R}{1 + \sqrt{\frac{A_i c_i^R}{B_i c_i^M}}} r_i, \quad \forall i \in \mathcal{A},$$

hence also:
$$s_i^M = \frac{c_i^M}{1 + \sqrt{\frac{B_i c_i^M}{A_i c_i^R}}} r_i, \quad \forall i \in \mathcal{A}.$$

Now, these results can be substituted in (P2d) to obtain:
$$\psi_i = -\frac{\left(\sqrt{\frac{A_i}{c_i^M}} + \sqrt{\frac{B_i}{c_i^R}}\right)^2}{E_i} r_i^{-1}, \quad \forall i \in \mathcal{A}.$$

□

The results of Proposition 3.3 allow to write a simplified version of problem (P2). Before showing it, we discuss the mentioned results.

Consider formulae (6a) and (6b): given a certain number of resources, for each user class it is possible to directly compute the optimal number of Map and Reduce slots to run at the maximum concurrency level allowed on the obtained $r_i$ VMs.

Formula (6c) is better described writing $\psi_i$ in terms of $h_i$. In this way, we have: $r_i = K_i h_i$, $\forall i \in \mathcal{A}$. Parameters $K_i$, therefore, are the minimum number of VMs needed to carry out exactly one job of class $i$ to meet the given deadline. On the other hand, given $r_i$ available VMs, it is possible to evaluate the maximum concurrency level that can be obtained. From this observation, formula (6c) can be rewritten to compute the number of resources needed to attain a specific concurrency level:

$$r_i^{up} = \frac{K_i}{\psi_i^{low}} = K_i H_i^{up}, \quad \forall i \in \mathcal{A}, \tag{8a}$$

$$r_i^{low} = \frac{K_i}{\psi_i^{up}} = K_i H_i^{low}, \quad \forall i \in \mathcal{A}. \tag{8b}$$

Formulae (8a) and (8b) define two new sets of parameters, $r_i^{up}$ and $r_i^{low}$, which are the optimal number of resources needed to complete jobs of class $i$ on time at maximum or minimum concurrency level, respectively. $r_i^{up}$ and $r_i^{low}$ appear in the reduced formulation of problem (P2) exploiting Proposition 3.3, which is presented here:

$$\min_{\mathbf{r}} \quad \sum_{i=1}^{N} \bar{\rho} r_i + \sum_{i=1}^{N} \left(\alpha_i \frac{K_i}{r_i} - \beta_i\right) \tag{P3a}$$

subject to:
$$\sum_{i=1}^{N} r_i \leq R, \tag{P3b}$$

$$r_i^{low} \leq r_i \leq r_i^{up}, \quad \forall i \in \mathcal{A}. \tag{P3c}$$

As can be seen, problem (P3) is smaller than (P2), with $2N+1$ constraints and $N$ variables. Moreover, due to the $r_i$ appearing at denominator in the penalty term, the problem remains nonlinear and convex.

However, problems (P2) and (P3) force to centralize on a single node data characterizing different application classes. This approach would have been natural in the first release of Hadoop where a single Job Tracker was in charge of both assigning resources to jobs and ensuring they were carried out. Instead, Hadoop 2.x distributes these duties among the RM and the AMs, so a more appealing approach would be to set them apart and solve distinct problems. In order to address this issue, we propose a distributed formulation where each CM and the RM solve smaller local problems. The decision variables are split among all these entities and the bargaining for resources becomes a game involving $N+1$ players.

## 4 Game Theoretic Formulation

Reformulating problem (P2) in a distributed way entails writing partial problems tailored to each entity taking part in resource allocation, i.e., the RM and CMs. Furthermore, we need to introduce mechanisms to take into account the penalties also in the RM partial problem, otherwise it would not consider the concurrency level and penalty consequences.

Given the reference scenario presented in Section 2 and Figure 1, we propose a game where, iteratively, the RM assigns VMs to the CMs, while they determine the concurrency level and the optimal distribution of slots for Map and Reduce tasks on the acquired resources. If the current assignment is not satisfactory, each CM can offer to pay a higher price for extra resources, then the RM assigns VMs to the best bidders.

This mechanism naturally leads to making $r_i$ as decision variables of the RM problem, whilst $s_i^M$, $s_i^R$, and $\psi_i$ are decision variables of the CM problems. Conversely, $s_i^M$, $s_i^R$, and $\psi_i$ are parameters for the RM problem, like $r_i$ for each CM problem.

Remarkably, the formulation that is to be laid out in this section belongs to the category of GNEPs. These are a generalization of classical Nash games where each player's strategy set is not fixed, but depends on the strategies adopted by the opponents. Indeed, the RM affects CMs' strategy sets by imposing an allowance of VMs and, in turn, CMs influence the RM with bids for resources.

In this section we introduce some more parameters and decision variables, summarized in Tables 3 and 4. Nonetheless, the parameters and variables already presented in Tables 1 and 2 are still of interest in the following models.

### 4.1 Class Managers

Class Managers are responsible for negotiating with the RM the assignment of a suitable number of VMs to carry out the submitted jobs. This behavior is modeled with a problem where $r_i$, the number of assigned VMs, are parameters, i.e., no CM can change their value. Therefore, the model features only constraints (P2c), (P2d), and (P2e), while its objective function takes into account only the penalty term of (P2a). Conversely, the RM problem might end up not being aware of the penalties applied to job rejections, then we introduce a new set of variables, $\rho_i^a$. These have the interpretation of bids placed by the CMs to virtually "buy" resources from the RM, but this aspect will be clarified in Sections 4.2 and 4.4. For the time being, it is sufficient to know that $\rho_i^a$ cannot be lower than the cost of execution $\bar{\rho}$, nor greater than some given upper bounds $\rho_i^{up}$.

Table 3: Distributed Model Parameters

| | |
|---|---|
| $\rho_i^{up}$ | Maximum bid CM $i$ can place to obtain resources |
| $\hat{\rho}$ | Maximum price the RM can set, $\max_i \rho_i^{up}$ |
| $r_i^{up}$ | Optimal assignment of VMs for job class $i$ executing at maximum concurrency |
| $r_i^{low}$ | Optimal assignment of VMs for job class $i$ executing at minimum concurrency |
| $p_i$ | Penalty paid for each VM not assigned to AM $i$ with respect to the optimal requirement for the execution at maximum concurrency |
| $\xi_i^M$ | Coefficient for evaluating the optimal number of Map slots to execute in job class $i$, given the assigned resources $r_i$ |
| $\xi_i^R$ | Coefficient for evaluating the optimal number of Reduce slots to execute in job class $i$, given the assigned resources $r_i$ |
| $K_i$ | Optimal number of VMs to carry out one job of class $i$ within the given deadline |

Table 4: Distributed Model Decision Variables

| | |
|---|---|
| $\rho_i^a$ | Bid placed by CM $i$ to obtain resources |
| $\rho$ | Price set by the RM for one hour of computation on one VM |
| $y_i$ | Logical variable, true when CM $i$ offers more than the price set by the RM for VMs |

In this model the decision variables are $\psi_i$, $\rho_i^a$, $s_i^M$, and $s_i^R$; for each application class $i \in \mathcal{A}$ they represent scalar variables. This means that each CM has to solve a small continuous convex problem, which can be solved quickly adopting state-of-the-art solvers. Additionally, the problem becomes so simple that the solution can be calculated in closed form.

The problem can be formulated as follows:

$$\min_{\psi_i, \rho_i^a, s_i^M, s_i^R} \quad \alpha_i \psi_i - \beta_i \quad \text{(P4a)}$$

subject to:

$$\bar{\rho} \leq \rho_i^a \leq \rho_i^{up}, \quad \text{(P4b)}$$

$$\psi_i^{low} \leq \psi_i \leq \psi_i^{up}, \quad \text{(P4c)}$$

$$\frac{A_i}{s_i^M \psi_i} + \frac{B_i}{s_i^R \psi_i} + E_i \leq 0, \quad \text{(P4d)}$$

$$\frac{s_i^M}{c_i^M} + \frac{s_i^R}{c_i^R} \leq r_i, \quad \text{(P4e)}$$

$$s_i^M \geq 0, \quad \text{(P4f)}$$

$$s_i^R \geq 0, \tag{P4g}$$
$$\psi_i \geq 0, \tag{P4h}$$
$$\rho_i^a \geq 0. \tag{P4i}$$

Recall that, in this class of problems, $r_i$ play the role of parameters, since they are not controlled by CMs. Thus, constraint (P4e) sets a bound on the possible values attained by $s_i^M$ and $s_i^R$, which in turn affect $\psi_i$. Due to this, the set of strategies each CM can play depends on the strategy adopted by the RM and is indirectly influenced by other CMs strategies.

## 4.2 Resource Manager

The Resource Manager receives from all the CMs requests for a certain number of VMs. Its role is to allocate the available resources, trying to satisfy all the requests to guarantee the timely completion of the jobs. Consequently, $r_i$ are the only decision variables. On the other hand, the model requires new variables: $\rho$ and $y_i$. Notice that, in this case, we have just one instance with variables subscripted over the whole set $\mathcal{A}$, contrasting to the formulation proposed in Section 4.1, where the model appears in $N$ instances separately solved by each CM.

The virtual pricing mechanism introduced to distribute the original problem is centered on the variables $\rho$ and $y_i$ and is inspired by the pricing policy enforced by Amazon for the allocation of on spot VMs. These are a class of VMs that can be acquired without previous agreements, contrasting to reserved VMs, but do not provide any availability guarantees to customers. Indeed, it is possible to apply for a number of such VMs offering to pay a unit price. Based on the received offers and the current availability of resources, Amazon sets a single on spot price. Now, customers who offered enough get their share of on spot VMs paying only the on spot price, which can be lower than their bids. However, Amazon may at any moment terminate them raising the on spot price to a value that is higher than the received bids.

In our setting, the on spot price is $\rho$ and is set by the RM. Every time the RM solves its problem, it determines the price based on the bids $\rho_i^a$ placed by the CMs. Every CM that offers more than $\rho$ is allowed to receive VMs up to the optimal upper bound $r_i^{up}$. On the other hand, CMs placing a bid lower than $\rho$ will receive exactly their minimum resource share, i.e., $r_i^{low}$. The parameters obtained in equations (8) provide sensible bounds to $r_i$, avoiding job starvation and waste of resources at once. Variables $y_i$ play their role in determining, through proper constraints, whether CM $i$ placed a bid greater than $\rho$ or not. In any case, the VMs unit price is $\rho$, for both overbidding and underbidding CMs. In this sense, CMs are not competing for resources *tout court*, but placing bids to obtain extra VMs over their minimum guaranteed allowance.

Clearly, to implement this behavior, it is necessary to use as objective function the total "revenue" and maximize it. However, in this way the model has no information about penalties. As long as the cluster capacity is not saturated, this is not much of an issue, since the tendency is to assign as many VMs as possible up to a value that cannot exceed $r_i^{up}$, the optimal requirement to have the maximum concurrency level. Two aspects are noteworthy. First, in this situation $r_i^{up}$ are obtained with an exact formula, exploiting the parameter that define the job profile, and this is the reason why VMs are not wasted for job classes that do not need more resources. Otherwise, the property would be recovered only with a fine calibration of the parameters. Secondly, in the general case where the cluster capacity is filled, thus making the CMs actually compete for resources, a model just considering this simple pricing mechanism would produce solutions with resources blindly spread among all the CMs, exposing cluster owners to avoidable penalties.

A more appealing objective function should have a term related to the job rejection penalties in addition to the one expressing the virtual revenue. Since the $\psi_i$ appear in this problem as parameters, such a term needs to depend on the assigned number of VMs, $r_i$, instead of the level of concurrency. Further, a similar penalty term should discourage the assignment of less resources than needed, but at the same time should not foster an uncontrolled growth of $r_i$ when there is no need to add more VMs. With these issues in mind, we introduce a new set of parameters, $p_i$, which are an equivalent penalty coefficient normalized on the lacking resources.

The complete model for the RM is the following:

$$\max_{\mathbf{r},\mathbf{y},\rho} \quad \sum_{i=1}^{N} (\rho - \bar{\rho}) r_i - \sum_{i=1}^{N} p_i (r_i^{up} - r_i) \tag{P5a}$$

subject to:

$$\sum_{i=1}^{N} r_i \leq R, \tag{P5b}$$

$$r_i \geq r_i^{low}, \quad \forall i \in \mathcal{A}, \tag{P5c}$$

$$r_i \leq (r_i^{up} - r_i^{low}) y_i + r_i^{low}, \quad \forall i \in \mathcal{A}, \tag{P5d}$$

$$\bar{\rho} \leq \rho \leq \hat{\rho}, \tag{P5e}$$

$$\rho - \rho_i^a \leq M(1 - y_i), \quad \forall i \in \mathcal{A}, \tag{P5f}$$

$$\rho_i^a - \rho \leq M y_i, \quad \forall i \in \mathcal{A}, \tag{P5g}$$

$$r_i \geq 0, \quad \forall i \in \mathcal{A}, \tag{P5h}$$

$$y_i \in \{0, 1\}, \quad \forall i \in \mathcal{A}, \tag{P5i}$$

$$\rho \geq 0. \tag{P5j}$$

In the objective function (P5a), we can see the second term expressing the penalty contribution. Instead, the first term quantifies the virtual gain obtained by the RM for the resources it assigns to CMs.

Constraints (P5c) and (P5d) are the lower and upper bounds on the resources assigned to each CM. In particular, the upper bound is $r_i^{up}$ when CM $i$ offers more than the price set by the RM, otherwise it is $r_i^{low}$, thus forcing $r_i = r_i^{low}$. This behavior is obtained through constraints (P5f) and (P5g), which are examples of the so called "big M" constraints; they make sure that $y_i$ has value 1 when CM $i$ offers enough, 0 otherwise. Now, if $\rho_i^a > \rho$, then $y_i = 1$ and the $i$-th constraint (P5d) reads $r_i \leq r_i^{up}$. Otherwise $y_i = 0$ and (P5d) becomes $r_i \leq r_i^{low}$, which holds with equality due to (P5c). For constraints (P5f) and (P5g) to achieve their intended goal, it must hold $M \geq \rho_i^{up}$, $\forall i \in \mathcal{A}$. Constraints (P5e) ensure that the price set by the RM is not less than the unit cost for running a VM in the cluster. Further, they avoid unrealistic solutions where the RM sets a price several orders of magnitude greater than the bids without assigning any extra resources, basically ignoring penalties. The parameter enforcing this behavior is $\hat{\rho}$, defined as $\hat{\rho} \triangleq \max_i \rho_i^{up}$. We also notice that $\hat{\rho}$ satisfies the properties required of $M$, hence they can be set equal to each other.

As in problem (P5) variables $r_i$ are taken as real nonnegative. However, in this case we are not solving the continuous relaxation of the problem, as variables $y_i$ are binary and they are treated as such. This means that model (P5) is a mixed integer nonlinear problem. In fact, even if constraint (P2d) is not part of this problem, the objective function involves the product $\rho r_i$.

### 4.3 Analysis

It is possible to further analyze problem (P4), as shown in the following. Since problem (P4) is convex and Slater constraint qualification holds, the KKT conditions are necessary and sufficient for optimality. The associated Lagrangian is:

$$
\begin{aligned}
\mathcal{L}(\psi_i, \rho_i^a, s_i^M, s_i^R) = {} & \alpha_i \psi_i - \beta_i + a_i \left(\bar{\rho} - \rho_i^a\right) + b_i \left(\rho_i^a - \rho_i^{up}\right) \\
& + c_i \left(\psi_i^{low} - \psi_i\right) + d_i \left(\psi_i - \psi_i^{up}\right) \\
& + e_i \left(\frac{A_i}{s_i^M \psi_i} + \frac{B_i}{s_i^R \psi_i} + E_i\right) \\
& + f_i \left(\frac{s_i^M}{c_i^M} + \frac{s_i^R}{c_i^R} - r_i\right) \\
& - g_i s_i^M - k_i s_i^R - l_i \psi_i - o_i \rho_i^a.
\end{aligned}
\tag{9}
$$

Hence, the KKT conditions are:

$$
\frac{\partial \mathcal{L}}{\partial \rho_i^a} = -a_i + b_i - o_i = 0, \tag{10a}
$$

$$
\frac{\partial \mathcal{L}}{\partial \psi_i} = \alpha_i - c_i + d_i - \frac{e_i A_i}{s_i^M \psi_i^2} - \frac{e_i B_i}{s_i^R \psi_i^2} - l_i = 0, \tag{10b}
$$

$$
\frac{\partial \mathcal{L}}{\partial s_i^M} = -\frac{e_i A_i}{\psi_i \left(s_i^M\right)^2} + \frac{f_i}{c_i^M} - g_i = 0, \tag{10c}
$$

$$
\frac{\partial \mathcal{L}}{\partial s_i^R} = -\frac{e_i B_i}{\psi_i \left(s_i^R\right)^2} + \frac{f_i}{c_i^R} - k_i = 0, \tag{10d}
$$

with the following complementary slackness conditions:

$$
a_i \left(\bar{\rho} - \rho_i^a\right) = 0, \quad a_i \geq 0, \tag{11a}
$$

$$
b_i \left(\rho_i^a - \rho_i^{up}\right) = 0, \quad b_i \geq 0, \tag{11b}
$$

$$
c_i \left(\psi_i^{low} - \psi_i\right) = 0, \quad c_i \geq 0, \tag{11c}
$$

$$
d_i \left(\psi_i - \psi_i^{up}\right) = 0, \quad d_i \geq 0, \tag{11d}
$$

$$
e_i \left(\frac{A_i}{s_i^M \psi_i} + \frac{B_i}{s_i^R \psi_i} + E_i\right) = 0, \quad e_i \geq 0, \tag{11e}
$$

$$f_i \left( \frac{s_i^M}{c_i^M} + \frac{s_i^R}{c_i^R} - r_i \right) = 0, \quad f_i \geq 0, \tag{11f}$$

$$g_i s_i^M = 0, \quad g_i \geq 0, \tag{11g}$$

$$k_i s_i^R = 0, \quad k_i \geq 0, \tag{11h}$$

$$l_i \psi_i = 0, \quad l_i \geq 0, \tag{11i}$$

$$o_i \rho_i^a = 0, \quad o_i \geq 0. \tag{11j}$$

We proceed similarly to the analysis carried out in Section 3, obtaining results analogous to those of Propositions 3.2 and 3.3.

**Proposition 4.1.** *The optimal solution to problem* (P4) *can be computed with the following formulae:*

$$s_i^M = \frac{c_i^M}{1 + \sqrt{\frac{B_i}{A_i} \frac{c_i^M}{c_i^R}}} r_i, \tag{12a}$$

$$s_i^R = \frac{c_i^R}{1 + \sqrt{\frac{A_i}{B_i} \frac{c_i^R}{c_i^M}}} r_i, \tag{12b}$$

$$\psi_i = -\frac{\left( \sqrt{\frac{A_i}{c_i^M}} + \sqrt{\frac{B_i}{c_i^R}} \right)^2}{E_i} r_i^{-1}. \tag{12c}$$

*Proof.* To begin with, observe that $g_i = k_i = l_i = o_i = 0$, since in every feasible solution all the variables take positive values.

Further, it can be easily shown that $\psi_i = \psi_i^{low}$ if and only if $r_i \geq r_i^{up}$. Indeed, the sets represented by constraints (P4d) and (P4e) are disjoint if $\psi_i = \psi_i^{low}$ and $r_i < r_i^{up}$, thus the feasible region is empty.

Now, we distinguish the cases $r_i = r_i^{up}$ and $r_i < r_i^{up}$. In the first case, formula (12c) yields $\psi_i = \psi_i^{low}$ due to the definition of $r_i^{up}$, equation (8a). Moreover formulae (12a) and (12b) give feasible values, hence we have an optimal solution.

In the second case, for our preliminary observation holds $\psi_i > \psi_i^{low}$, then from (11c) descends $c_i = 0$. We can then prove that $e_i > 0$, according to (10b), and $f_i > 0$, due to (10c): this means that constraints (P4d) and (P4e) are active in every optimal solution. Using equations (10c) and (10d) we get:

$$s_i^M = s_i^R \sqrt{\frac{A_i c_i^M}{B_i c_i^R}}.$$

Substituting into (P4d) and (P4e) we obtain formulae (12). □

The coefficients multiplying $r_i$ or their reciprocal are constants depending on job profiles, so they can be computed once and for all, reducing the solving procedure of problem (P4) to an algebraic update of just three closed formulae per CM. Moreover, notice that the results of Proposition 4.1 are formally the same obtained in Proposition 3.3.

Notice that Proposition 4.1 provides the only optimal solution in closed form when $\psi_i \in \left( \psi_i^{low}, \psi_i^{up} \right]$. On the other hand, objective (P4a) does not consider any term pushing towards an efficient allocation of resources, hence any configuration guaranteeing maximum level of concurrency is equivalently optimal for problem (P4). Even losing the uniqueness property, we can adopt the results of Proposition 4.1 also when $\psi_i = \psi_i^{low}$ to determine the most efficient among optimal solutions at the maximum level of concurrency, recovering the efficiency property guaranteed by the preliminary, centralized problem (P2).

**Algorithm 4.1** Best Reply
---
1: $r_i \leftarrow r_i^{low}, \forall i \in \mathcal{A}$
2: $s_i^M \leftarrow s_i^{M,low}, \forall i \in \mathcal{A}$
3: $s_i^R \leftarrow s_i^{R,low}, \forall i \in \mathcal{A}$
4: $\psi_i \leftarrow \psi_i^{up}, \forall i \in \mathcal{A}$
5: $\rho_i^a \leftarrow \bar{\rho}, \forall i \in \mathcal{A}$
6: **repeat**
7:     $r_i^{old} \leftarrow r_i, \forall i \in \mathcal{A}$
8:     RM solves problem (P5)
9:     **for all** $i \in \mathcal{A}$ **do**
10:         CM $i$ solves problem (P4)
11:         **if** $\psi_i > \psi_i^{low}$ **then**
12:             $\rho_i^a \leftarrow \max\{\rho_i^a, \rho\} + \lambda \rho_i^{up}$
13:         **end if**
14:     **end for**
15:     $\varepsilon \leftarrow \sum_{i=1}^{N} \frac{|r_i - r_i^{old}|}{r_i^{old}}$
16: **until** $\varepsilon < \bar{\varepsilon}$
---

Problem (P5) does not lend itself to such exact results as the ones presented in this section, hence we will face it with ordinary mathematical programming techniques.

## 4.4 Iterative Approach

According to the considerations presented in Section 4.1, in this game all the CMs have strategy sets that are functions of the strategy adopted by the RM. Similarly, the strategy set of the RM depends on the bids, $\rho_i^a$, offered by the CMs. As previously stated, this makes the game an example of GNEP. In order to have the system converge to an equilibrium, problems (P4) and (P5) are solved iteratively until a stopping criterion is fulfilled.

Algorithm 4.1 starts assigning the initial values to all the decision variables. Then it starts a loop performing the actual iterative procedure to solve the game. First of all, the current configuration is saved in a set of auxiliary variables, in order to compare it with the updated solution at the end of each iteration and check the stopping criterion. Here we propose to stop the computation when the total relative increment is below a given tolerance. At line 8 the RM solves its problem, setting new values for $r_i$ and making them available to every CM. In the following nested loop, each CM solves its instance of problem (P4) and, possibly, places a higher bid for extra resources. Condition $\psi_i > \psi_i^{low}$ at line 11 checks whether CM $i$ is rejecting any jobs: in this case, it places a higher bid trying to obtain a greater share of resources, so as to reduce the penalty-related expenditure. Parameter $\lambda$ is a fraction of the maximum possible offer added to either the previous bid or the price set by the RM, hence $\lambda \in (0,1)$.

Notice that Algorithm 4.1 is meant to be solved in a distributed fashion, with the loop at line 9 executed in parallel, one iteration per CM. The algorithm execution requires that the RM sends $r_i$ to each CM and the latter sends back the updated bids $\rho_i^a$. In this way, the RM does not need any sensitive knowledge on the SLAs.

## 4.5 Integer Solution Heuristic

We have previously observed that, since the continuous relaxations of the proposed models are considered for performance reasons, we need to enforce the integrality of variables $r_i$, $s_i^M$, and $s_i^R$ to obtain a solution useful in practice. In order to fulfill this necessity, we adopt a heuristic approach, which is presented in this section. The algorithm consists of two main phases: initially, the input data, i.e., the characterizing parameters for cluster performance and those related to Quality of Service (QoS) and SLAs, are fed to a continuous formulation; afterwards, the continuous solution is transformed in an integer one with the steps outlined in the following. We propose a heuristic that does not depend on how the first phase is performed, hence this approach applies both to the centralized and the distributed formulations previously discussed. Algorithm 4.2 shows pseudo-code for the proposed rounding heuristic.

Constraints (P4d) are approximate formulae, hence can be relaxed when looking for nearly optimal integer solutions. Moreover, the constraint formulation proposed in Section 2 does not provide strong theoretical guarantees that deadlines are met, even without the latest approximation. Building upon this observation, in Algorithm 4.2 we propose a viable heuristic to obtain, given the optimal real solution provided by either problem (P2) or Algorithm 4.1, an integer solution that is feasible with respect to all constraints but (P4d). Note that the RM must do so before starting VMs, since it is not possible to launch a portion of a VM. In the following, we denote with $\left(\hat{\mathbf{r}}, \hat{\boldsymbol{\psi}}, \hat{\mathbf{s}}^M, \hat{\mathbf{s}}^R\right)$ either the optimal solution of the continuous relaxation of our centralized resource allocation problem (P2), or the game equilibrium obtained through Algorithm 4.1.

We can easily prove that the loop starting at line 3 is enough to find a set of values of the $r_i$ variables satisfying the following constraint:

$$\sum_{i=1}^{N} r_i \leq R. \qquad (13)$$

**Proposition 4.2.** *Constraint* (13) *can be satisfied decrementing, at most, all the $r_i$ once.*

*Proof.* Let $\left(\hat{\mathbf{r}}, \hat{\boldsymbol{\psi}}, \hat{\mathbf{s}}^M, \hat{\mathbf{s}}^R\right)$ be a continuous optimal solution. Then, suppose to decrement all the $r_i$ once: it follows $r_i = \lfloor \hat{r}_i \rfloor, \forall i \in \mathcal{A}$. Now, $\lfloor \hat{r}_i \rfloor \leq \hat{r}_i, \forall i \in \mathcal{A}$, hence:

$$\sum_{i=1}^{N} r_i \leq \sum_{i=1}^{N} \hat{r}_i \leq R. \quad \square$$

**Algorithm 4.2** Solution Rounding
───────────────────────────────────────────
1: sort $\mathcal{A}$ according to increasing $\alpha_i$
2: $r_i \leftarrow \lceil \hat{r}_i \rceil, \ \forall i \in \mathcal{A}$
3: **for all** $j \in \mathcal{A}$ **do**
4:     **if** $\sum_{i=1}^{N} r_i > R$ **then**
5:         $r_j \leftarrow r_j - 1$
6:     **end if**
7: **end for**
8: $s_i^M \leftarrow \lceil \hat{s}_i^M \rceil, \ \forall i \in \mathcal{A}$
9: $s_i^R \leftarrow \lceil \hat{s}_i^R \rceil, \ \forall i \in \mathcal{A}$
10: **for all** $j \in \mathcal{A}$ **do**
11:     **while** $s_j^M/c_j^M + s_j^R/c_j^R > r_j$ **do**
12:         $s_j^R \leftarrow s_j^R - 1$
13:         **if** $s_j^M/c_j^M + s_j^R/c_j^R > r_j$ **then**
14:             $s_j^M \leftarrow s_j^M - 1$
15:         **end if**
16:     **end while**
17: **end for**
───────────────────────────────────────────

Moreover, the nested loop at line 11 has complexity $\mathcal{O}(1)$ in the worst case. Indeed, we can prove that the number of operations it requires does not depend on $N$.

**Proposition 4.3.** *Each of the following constraints:*

$$\frac{s_i^M}{c_i^M} + \frac{s_i^R}{c_i^R} \leq r_i, \quad \forall i \in \mathcal{A}, \tag{14}$$

*can be satisfied in no more than $\omega_i + 1$ iterations, where $\omega_i = \min\{c_i^M, c_i^R\}$.*

*Proof.* Let $\left(\hat{\mathbf{r}}, \hat{\boldsymbol{\psi}}, \hat{\mathbf{s}}^M, \hat{\mathbf{s}}^R\right)$ be a continuous optimal solution. Consider $i \in \mathcal{A}$. Since $\lceil \hat{s}_i^M \rceil - 1 \leq \hat{s}_i^M$ and $\lceil \hat{s}_i^R \rceil - 1 \leq \hat{s}_i^R$, it holds:

$$\frac{\lceil \hat{s}_i^M \rceil - 1}{c_i^M} + \frac{\lceil \hat{s}_i^R \rceil - 1}{c_i^R} \leq \hat{r}_i.$$

Now, after $\omega_i + 1$ iterations, we obtain:

$$\frac{\lceil \hat{s}_i^M \rceil - (\omega_i + 1)}{c_i^M} + \frac{\lceil \hat{s}_i^R \rceil - (\omega_i + 1)}{c_i^R} < \hat{r}_i - 1 \leq \lfloor \hat{r}_i \rfloor.$$

Hence, using the loop shown in Algorithm 4.2 at line 11, it is possible to satisfy the $i$-th constraint (14) with at most $\omega_i + 1$ iterations even in the worst case, when $r_i = \lfloor \hat{r}_i \rfloor$. □

Given that both the main loops are linear in $N$, the complexity of Algorithm 4.2 is dominated by the adopted sorting algorithm, i.e., it is $\mathcal{O}(N \log N)$. Furthermore, notice that Algorithm 4.2 can be easily parallelized. In the distributed scenario, from line 8 every CMs can separately round their variables at the same time.

# 5 Results

In this section we present and examine experimental results obtained by applying the solution methods proposed in Sections 3 and 4. We implemented and evaluated two different models: problem (P2), named *centralized* in the following, and Algorithm 4.1 where problem (P4) is solved using the results of Proposition 4.1, from now on *distributed*. Both, after completing, provide their results as input for Algorithm 4.2.

Since problem (P2) is the natural extension of the model studied and validated in [13] to a new setting with capacity constraints, the centralized algorithm is used as base case. Further, building upon the results shown in [21], we assume the validity of the underlying performance model.

The models have been implemented using the mathematical programming language AMPL [22] and solved using Knitro 9.0.1 [23]. All the analyses have been run on an Ubuntu 14.04 VM featuring 14 GB RAM hosted on an Intel Xeon E5530 2.40 GHz CPU.

## 5.1 Design of Experiments

The analyses in this section intend to be representative of real Hadoop clusters. In order to do so, the experiments have been performed considering realistic job profiles extracted from MapReduce execution logs, as in [24], where the authors report profiles obtained for batch jobs. The associated deadlines are extracted from a uniform distribution in the range from 15 to 25 minutes.

After determining meaningful ranges for the parameters, the experiments have been executed on random instances obtained using uniform distributions, within the ranges reported in Table 5. Moreover, all the parameters computed from others are summarized in Table 6. Unless differently specified, all the problem instances of a given size are solved ten times with a fixed set of seeds for the pseudo-random number generator. In this way we achieve both repeatability of the experiments and meaningful, average results. Moreover, the different models are executed against the same data, making the outcomes comparable.

Parameter $\bar{\rho}$ takes into account three main contributions: the energy cost related to the operation of the physical servers hosting the VMs, the overhead related to server rooms cooling, and the price of physical servers. The unit energy cost considers European energy prices [25], [26]. Further, the power consumption is estimated taking into account reference benchmarks for servers in production environment, precisely SPECpower [27]. With them, the cost of one hour of computation of a single core, $\mathcal{E}$, is estimated. The cooling overhead is modeled applying a multiplier, the *power usage effectiveness* (PUE) [28], to the unit cost of energy. This coefficient is a metric to quantify the relevance of costs that are not directly imputable to the IT equipment in a data center, like lighting, cooling, etc. The price of servers is associated to the modeled computational units with a straight line depreciation method, dividing prices of representative servers, as can be found on manufacturers web sites, by the number of hours in the service life, assumed to be of four years, and the number of physical cores, obtaining the unit cost $\mathcal{S}$.

After estimating the three mentioned contributes, they are aggregated in $\bar{\rho}$. To map virtual CPUs on physical cores, a further reference benchmark, namely SPECvirt [29], was studied to understand the typical density, $d$, in produc-tion clusters. Here we define the density as the virtual to physical core ratio. Moreover, a reference VM class has been chosen, thus identifying the number of virtual cores per instance, $v$. In this paper we considered Amazon EC2 m3.large instances, general purpose VMs guaranteeing satisfying performance. With all these data, the formula for $\bar{\rho}$ reads:

$$\bar{\rho} = (\text{PUE} \cdot \mathcal{E} + \mathcal{S}) \frac{v}{d}. \tag{15}$$

Table 5: Parameters Uniform Distributions

| Parameter | Range | Units of Measurement |
|---|---|---|
| $\rho_i^{up}$ | $[5, 20]$ | [€ cents] |
| $H_i^{up}$ | $[5, 20]$ | [-] |
| $c_i^M$ | $[1, 4]$ | [-] |
| $c_i^R$ | $[1, 4]$ | [-] |
| $m_i$ | $[15000, 30000]$ | [€ cents] |
| $n_i^M$ | $[70, 1120]$ | [-] |
| $n_i^R$ | $[64, 64]$ | [-] |
| $M_i^{max}$ | $[16, 120]$ | [s] |
| $R_i^{max}$ | $[15, 75]$ | [s] |
| $Sh_{1,i}^{max}$ | $[10, 30]$ | [s] |
| $Sh_{typ,i}^{max}$ | $[30, 150]$ | [s] |
| $D_i$ | $[900, 1500]$ | [s] |
| $v$ | $[2, 2]$ | [-] |
| $d$ | $[3, 5]$ | [-] |
| PUE | $[1.2, 2.2]$ | [-] |
| $\mathcal{E}$ | $[0.06009, 0.06690]$ | [€ cents] |
| $\mathcal{S}$ | $[2.0615, 2.0615]$ | [€ cents] |

Table 6: Derived Parameters

| Parameter | Range | Units of Measurement |
|---|---|---|
| $H_i^{low}$ | $[4, 16]$ | [-] |
| $\psi_i^{low}$ | $[0.05, 0.2]$ | [-] |
| $\psi_i^{up}$ | $[0.0625, 0.25]$ | [-] |
| $M_i^{avg}$ | $[12.8, 96]$ | [s] |
| $R_i^{avg}$ | $[12, 60]$ | [s] |
| $Sh_{typ,i}^{avg}$ | $[24, 120]$ | [s] |
| $A_i$ | $[656, 107488]$ | [s] |
| $B_i$ | $[1854, 11430]$ | [s] |
| $C_i$ | $[132, 720]$ | [s] |
| $E_i$ | $[-1368, -180]$ | [s] |
| $\bar{\rho}$ | $[0.85344, 1.47246]$ | [€ cents] |
| $\alpha_i$ | $[300000, 9600000]$ | [€ cents] |
| $\beta_i$ | $[60000, 480000]$ | [€ cents] |
| $p_i$ | $[14.284, 34812]$ | [€ cents] |
| $\xi_i^M$ | $[0.19327, 3.53565]$ | [-] |
| $\xi_i^R$ | $[0.11609, 3.22693]$ | [-] |
| $K_i$ | $[0.86178, 1050.1]$ | [-] |
| $r_i^{up}$ | $[4.3089, 21002]$ | [-] |
| $r_i^{low}$ | $[3.4471, 16802]$ | [-] |

Other relevant parameters are those governing the penalties for job rejection, $\alpha_i$ and $\beta_i$. Running the centralized model (P2) with $\psi_i^{up} = \psi_i^{low}$, so that rejection is not possible, one can determine an average job cost. Penalties can reasonably be a couple of orders of magnitude greater than the job cost, hence we take $m_i$, the penalty associated to the rejection of one job in class $i$, as 100 times the average job cost. Now it is possible to set the penalty terms appearing in the objective functions (P2a) and (P4a) equal to zero at maximum concurrency level, when no penalties are paid, and to the value computed with the above mentioned parameters at minimum concurrency. For convenience we write $\psi_i^{up}$ and $\psi_i^{low}$ in terms of $H_i^{low}$ and $H_i^{up}$, thus getting the systems:

$$\begin{cases} \frac{\alpha_i}{H_i^{up}} - \beta_i = 0 \\ \frac{\alpha_i}{H_i^{low}} - \beta_i = m_i \left( H_i^{up} - H_i^{low} \right) \end{cases}, \quad \forall i \in \mathcal{A}, \tag{16}$$

that yield:

$$\alpha_i = m_i H_i^{up} H_i^{low}, \quad \forall i \in \mathcal{A}, \tag{17a}$$
$$\beta_i = m_i H_i^{low}, \quad \forall i \in \mathcal{A}. \tag{17b}$$

Owing to the interpretation of parameters $K_i$, discussed in Proposition 3.3, it is easy to obtain the penalty term normalized on lacking resources, $p_i$. Indeed, it is enough to divide the penalty value per job, $m_i$, by the resources requirement of each job, $K_i$. Hence, we get the formulae:

$$p_i = \frac{m_i}{K_i}, \quad \forall i \in \mathcal{A}. \tag{18}$$

In the end, we experimentally set the tolerance on the relative increment of $r_i$, appearing as stopping criterion in Algorithm 4.1, $\bar{\varepsilon} = 3\%$. Furthermore, Section 5.4 reports a sensitivity analysis to $\bar{\varepsilon}$. Differently, we set in all the experiments $\lambda = 0.05$ as fraction of the maximum possible offer to add when raising the bid.

## 5.2 Scenario-Based Analysis

In this section we discuss some preliminary analyses performed to verify whether our formulations exhibit behaviors we intuitively expect from the modeled applications. Analyses have been run with both the previously mentioned solution approaches, considering 100 and 1,000 CMs.

In order to compare the outcomes of the solution methods, in this section we do not present results averaged on a number of random instances. Instead, we randomly generated a dataset of size 1,000, then we shrank it as needed for smaller instances, adapting capacity to the different scale.

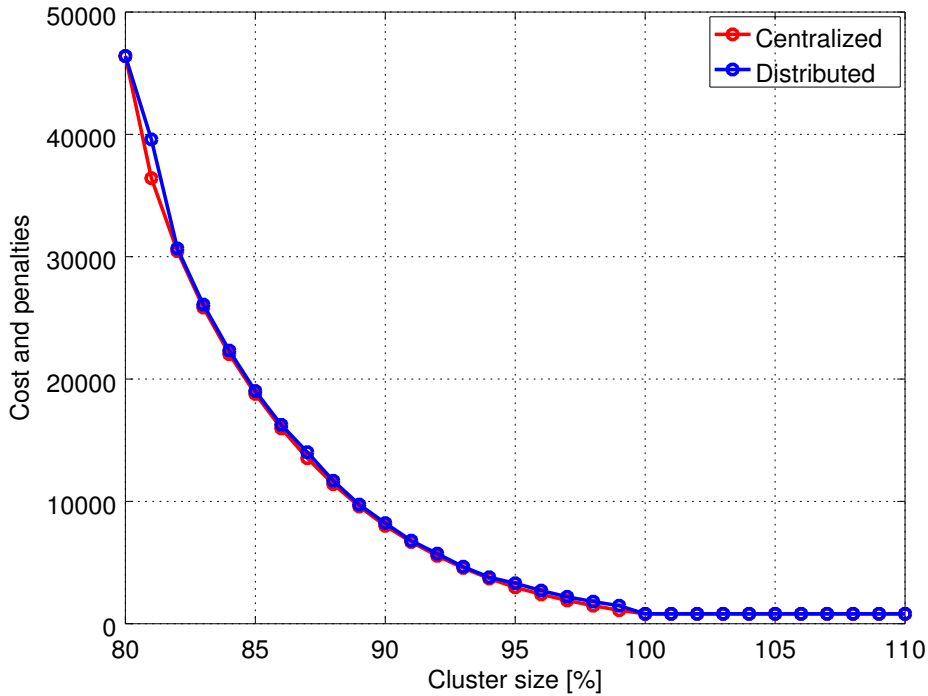

Figure 2: Decreasing capacity, 100 CMs

In the MapReduce paradigm we have two main dimensions governing performance: namely, resource capacity and deadlines. Isolating each of them allows to understand how the system will react to changes in one single aspect.

### 5.2.1 Decreasing Capacity

To begin with, we fix the deadlines $D_i$ and we vary the cluster capacity $R$. Now, starting with excess capacity, we expect to see constant running costs until the aggregate resources requirement for execution at the maximum concurrency level and the cluster capacity are comparable. If capacity keeps decreasing, at first some jobs are rejected, leading to penalties, then it falls below the minimum aggregate requirement and the problem becomes infeasible.

Notice that in this experiment we initially compute the optimal cluster size summing up all the $r_i^{up}$ and we set $R^o = \sum_{i=1}^{N} r_i^{up}$. Then we apply our models to instances with decreasing capacity, starting from $R = 1.1 R^o$. Figures 2 and 3 show slight shifts away from the optimal solution of the centralized problem, concentrated in proximity of the lowest feasible capacity levels.

### 5.2.2 Decreasing Deadlines

Alternatively, it is possible to fix the cluster capacity, $R$, decreasing deadlines and making them tighter. In this case, parameters $D_i$ act in nonlinear constraints, so we can predict an overall behavior, but the precise relation governing costs is less obvious. In the beginning, the reduction of the deadlines leads to an increase of the running costs, due to the allocation of more VMs to reduce execution times and meet tighter constraints. Then we expect a phase where the increment of costs accelerates, as jobs start to be rejected and penalties to be paid. In the end, the deadlines become so strict in relation to the available resources to not allow even minimal operation, hence the problem turns out to be infeasible.

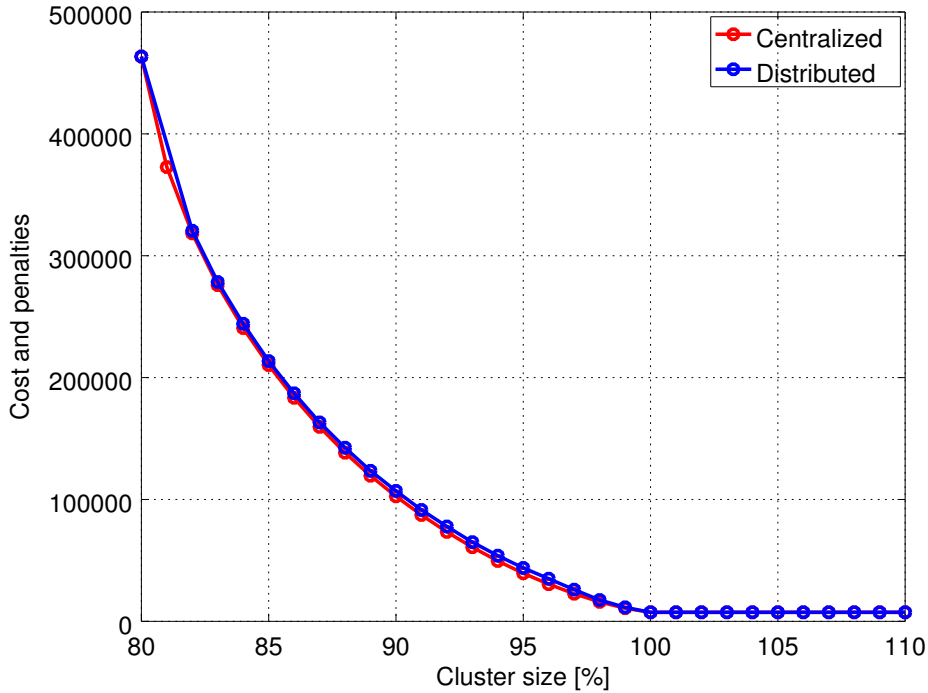

Figure 3: Decreasing capacity, 1000 CMs

In this experiment, we initially set the starting values for parameters $D_i$ and compute the corresponding optimal capacity, $R^o = \sum_{i=1}^{N} r_i^{up}$. Then we set $R = 1.1 R^o$ to start with excess capacity and gradually reduce $D_i$ by a percentage. Figures 4 and 5 highlight a great accordance with the expected behavior.

### 5.3 Scalability Analysis

After verifying that our models satisfy some basic properties we intuitively expect, and in line with the results published in [13], we proceed with a scalability analysis. Our goal is to verify that the proposed approach allows for solving the joint admission control and capacity allocation problem at runtime, to support cluster management. Moreover, we want to check that this is feasible even in realistic configurations with hundreds of different job classes.

We progressively increase the number of CMs from 20 to 500 with step 20, performing at every step ten runs with different extractions of the randomly generated parameters, then we compute the average results. Concerning total costs and penalties, we compare objective function (P2a), obtained with the central-ized, baseline approach and defined $\mathcal{C}^c$, with the sum of all the objectives (P4a) and the aggregate cost of energy, $\sum_{i=1}^{N} \bar{\rho} r_i$, of the solutions determined with Algorithm 4.1, dubbed $\mathcal{C}^d$. In this way we obtain from both the alternative solution methods the same information summarizing running costs and penalties. In Figure 6, we see that both the alternative solution methods, despite a quite high tolerance, $\bar{\varepsilon} = 3\%$, yield almost equivalent solutions in terms of aggregate execution costs and penalties.

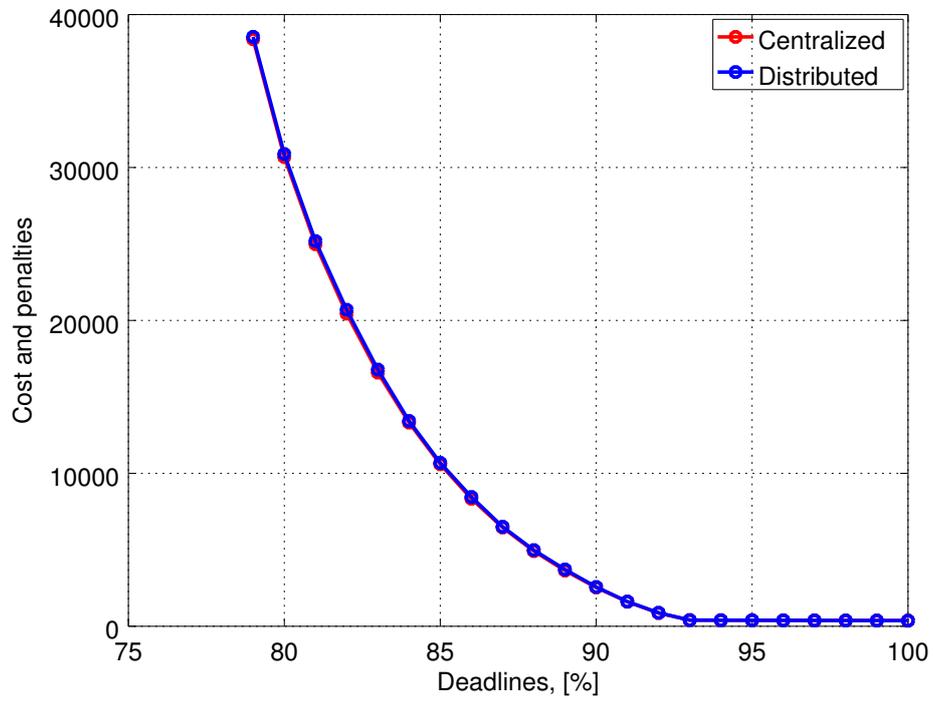

Figure 4: Decreasing deadlines, 100 CMs

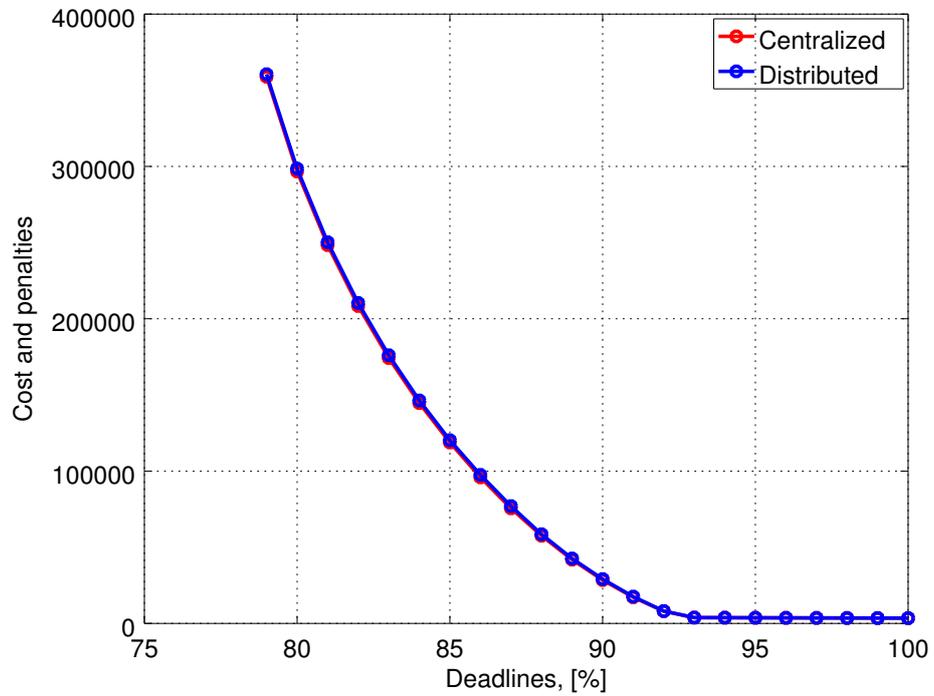

Figure 5: Decreasing deadlines, 1000 CMs

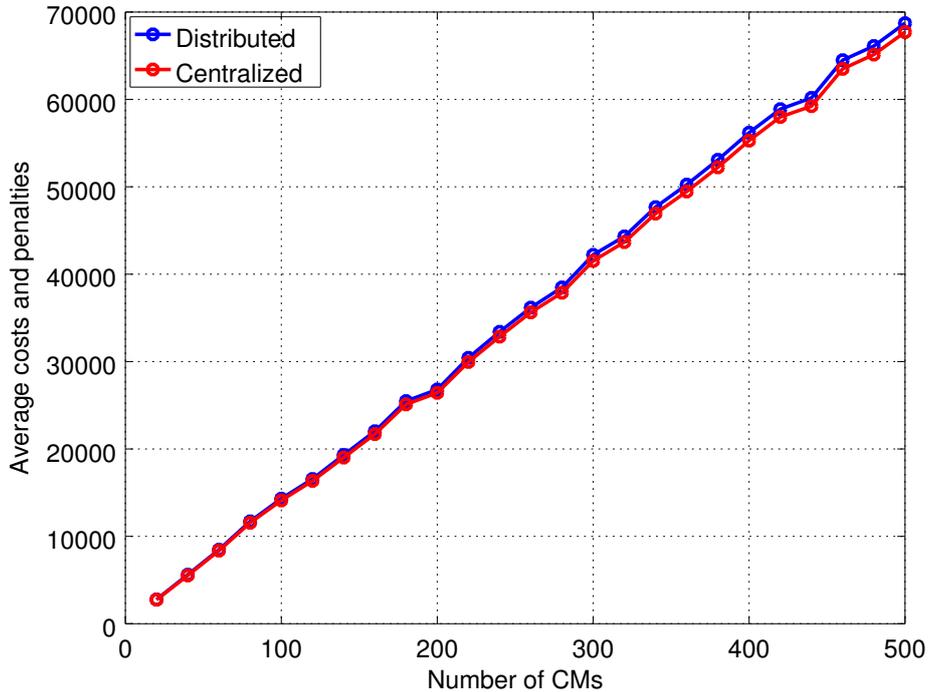

Figure 6: Costs, scalability analysis

Concerning solving times, instead, we instrumented our code to measure the timings associated to the solution algorithms. Further, whilst in the centralized approach this is enough to obtain relevant execution times, we should consider that in our testing environment Algorithm 4.1 runs on a single machine as serial code. In order to have a first approximate estimate of the actual execution times in a distributed environment, we divided the serially obtained timing by the number of CMs included in the instance at hand. This is justified as, in every iteration of Algorithm 4.1, the loop at line 9 would be executed concurrently by all the CMs at once in a distributed system. Moreover, we take into account the network-related delays by adding a term proportional to the number of iterations needed to reach convergence. Network time has been obtained by writing a micro-benchmark that sends through sockets two floats, in our system the parameters $\rho$ and $r_i$ coming from the RM, or $\rho_i^a$ coming from the CMs. The benchmark was run 100 times in a 100 Mb/s network. The aggregate results are reported in Figure 7, where it is evident that the distributed approach we propose scales better than the centralized one.

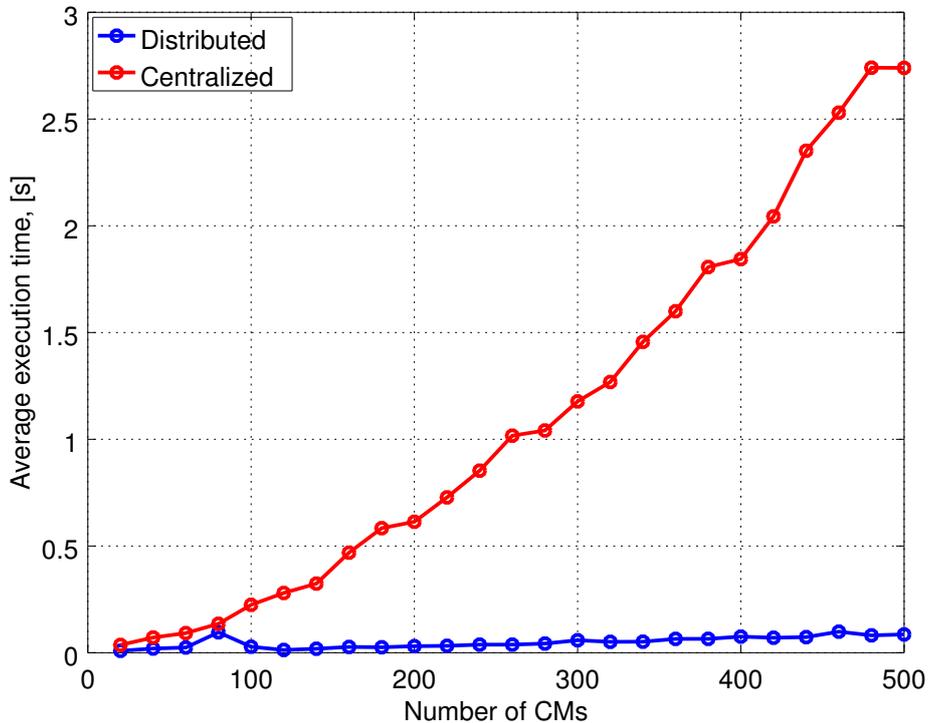

Figure 7: Solving time, scalability analysis

### 5.4 Stopping Criterion Tolerance Analysis

In this subsection we discuss how much the proposed solution method is sensitive to changes of the stopping criterion tolerance, $\bar{\varepsilon}$, in Algorithm 4.1. In order to do so, we repeated the experimental campaign described in Section 5.3 four times: first of all we consider $\bar{\varepsilon} = 3\%$ as for the already seen results, then we run the distributed algorithm with $\bar{\varepsilon} = 1\%$, $\bar{\varepsilon} = 5\%$, and $\bar{\varepsilon} = 10\%$.

Figure 8 reports the relative error on average total costs and penalties compared to the centralized solution, i.e., the following quantity:

$$\chi = \frac{\mathcal{C}^d - \mathcal{C}^c}{\mathcal{C}^c}, \qquad (19)$$

where $\mathcal{C}^d$ is the average total cost obtained with the distributed approach, while $\mathcal{C}^c$ is the average total cost yielded by the centralized method. We clearly see that all the different tolerance values overlap and relative errors do not exceed 2%. Hence, the proposed distributed approach has a very low sensitivity to the chosen stopping criterion tolerance.

## 6 Related Work

Nowadays, Hadoop is widely adopted in the ICT industry, often supporting core business activities. Hence, it is of paramount importance for users running

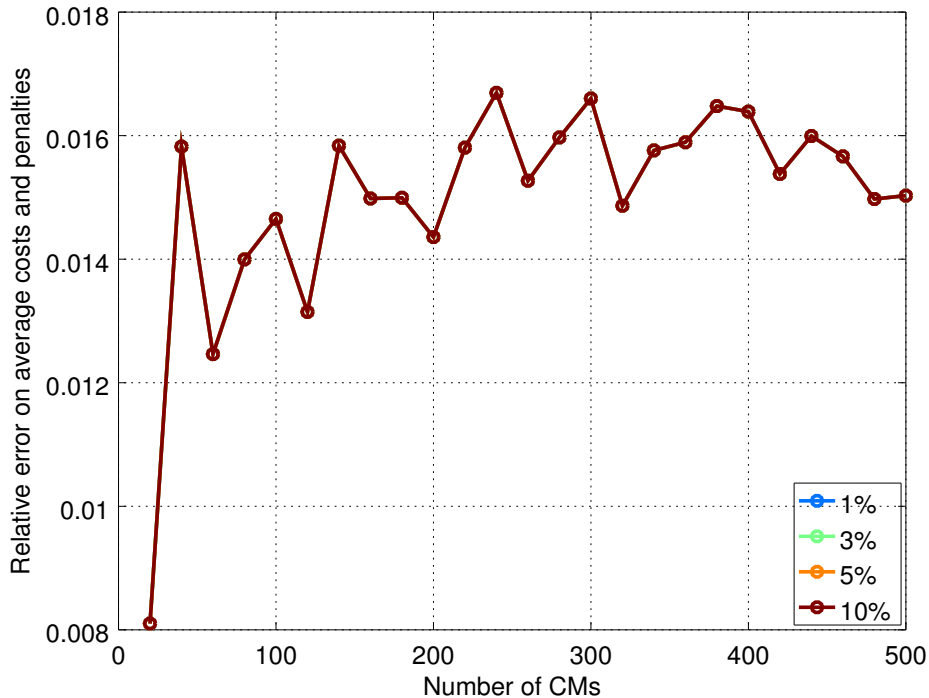

Figure 8: Sensitivity analysis

MapReduce applications to know in advance the job completion time. Technical literature focuses on two main points of interest: on one hand, there are various works addressing methods for the estimation of job completion times and other performance metrics; on the other hand, several publications present novel approaches to resource allocation and scheduling.

By evaluating a model of the Hadoop environment, it is possible to obtain an estimation of the time required for job completion. Two main approaches have been explored: simulation-based models implement the single constituents of Hadoop and of the job, replaying in a simulated environment the steps and delays of the real system; analytical models, on the other hand, define a mathematical representation of those constituents, avoiding the costs of running multiple simulations. Both approaches make use of information such as input dataset size, cluster resources, and Hadoop specific parameters. The computational effort and the time spent in running simulation-based models make them hardly fit for the purposes of runtime cluster management: thus, we hereby consider only analytical models, according to the focus of our paper.

The authors of [5] propose the Automatic Resource Inference and Allocation (ARIA) framework for estimating the makespan of jobs in MapReduce clusters. This approach relies on information extracted from the logs of previous executions of similar jobs. Adopting scheduling techniques, the authors prove lower and upper bounds on makespans. From these results, they derive formulae for performance prediction. They obtain both a conservative estimate, suitable for hard deadlines, and an alternative that does not offer guarantees of meeting deadlines, but boasts a relative error below 10% in the validation against measured timings.

The same research group went on in defining a performance model for Hadoop clusters [30]. At first they classify the job data processing phases into two categories, dividing common procedures carried out by the Hadoop framework from user-supplied code in Map and Reduce functions. Both are profiled, respectively with micro-benchmarks and extracting information from past executions logs, thus allowing for the estimation of job execution time when dealing with input of different sizes. In comparison to the measured values on a 66-node Hadoop cluster, the relative error on the predicted completion times ranges between 10% and 17%.

A more detailed model was developed in [31]. The CPU, network, and I/O cost is computed for each MapReduce phase as a function of different parameters specific of both the phase and the input data. In this technical report, neither validation analyses nor tests are reported. Moreover, none of the costs takes into account resource contention.

The model presented in [32] considers the execution time of each step in the Hadoop framework, modeling both computation and resource contention. Although this analytical model does not perform very well in reporting the real job execution time, the trend obtained when increasing the number of Map and Reduce tasks is quite reliable and can help in adjusting the number of tasks towards an optimal value.

Another performance model estimating the execution time by considering the single costs of the various phases of a MapReduce job is described in [33]. In this work, the authors go down to the very low level elements that determine the cost of single job phases, writing a 37-parameter model that provides execution times within 10% of those measured in a real cluster. Even with such an accurate model, the validation considers just single job executions.

In [34], the authors apply linear regression to several executions in real clusters, both physical and Cloud-based. In this study, the execution time of a job is a function of the input size and number of worker nodes. The experimental results on the physical cluster report a maximum error of 14%, while on the Cloud-based one the maximum error is 35%.

Vianna et al. [35] combine a precedence tree, which captures the precedence constraints among the different tasks of the same job, and a closed queueing network (QN), to reflect resource contention within the system. Solving this model, they obtain the final average job response time together with other performance metrics. The proposed method is validated through QN simulations and runs on an actual cluster, obtaining a deviation from real setup of less than 15%, but without considering multiple concurrent jobs.

Another approach is provided in [36], where the authors propose a performance evaluation method for Mean Value Analysis to solve a closed QN model, describing resource contention among tasks. The experimental results show that the increase in the number of slots in a MapReduce environment brings benefit until the resource contention becomes relevant and the tasks execution times start to increase as well. However, the model does not seem to adequately capture this behavior as the number of slots increases.

The models presented before may possibly be adopted for resource allocation and admission control purposes, for instance exploiting game-theoretic or optimization techniques. However, to the best of our knowledge, in literature there exist only a few examples of papers focusing on resource allocation, even less if we consider distributed approaches. In the following, we show some works that inquiry resource management and scheduling issues.

In [37], the authors propose an optimized approach to sharing clusters among MapReduce and Cloud applications. Relying on a Nash bargaining game, the authors develop modules for fair resource allocation and automatic VM migration, thus increasing resource utility and guaranteeing performance. They conduct validation on an 8-node server cluster, comparing the traditional method and the proposed hybrid environment, which outperforms the former in all the presented experiments.

Song et al. [38] put their attention on the scheduling of MapReduce jobs. In order to solve some issues of the standard schedulers supplied with Hadoop, the proposal is a novel approach that splits job scheduling and task scheduling. Both tasks are addressed with game-theoretic techniques. The proposed approach is validated through simulations, showing an improvement in both phases with respect to the FIFO scheduler.

Sandhom and Lai [39] address MapReduce optimization relying on a strong theoretical and statistical background. The authors highlight how allocating resources fairly can lead to decay in performance and efficiency: according to their proposal, users receive a budget and the system allocates resources according to the relevance of each users' current spending rate with respect to total demand, updating the shares in real time. The approach is validated with experiments on a real cluster, showing improvements upon a basic, fair share strategy.

Contrasting to the presented literature, our paper couples performance prediction and resource allocation, thus building a theoretical framework capable of managing cluster resources cost-effectively, while also guaranteeing, with a certain confidence, that SLAs are met.

# 7 Conclusions

In this paper we investigated the problem of resource allocation of MapReduce applications running on Hadoop clusters. In particular, we provided a scalable distributed approach for solving the joint admission control and capacity allocation problem based on a performance model available in technical literature.

Building upon the outcomes of this work, it is possible to investigate further open issues and relevant research questions. A project that is attracting increasing attention in the industry is Apache Tez [40]. It can be seen as the natural evolution of Hadoop, where workflows are not fixed anymore: the framework abstracts the dependency relationships among input, output, and intermediate data with Directed Acyclic Graphs (DAGs). An interesting development of this work is the extension of the model to consider the mechanisms governing Apache Tez, hence adapting our joint admission control and capacity allocation problem to the execution of complex DAGs.

In the end, we should be aware that the approximate formulae we use to estimate performance might incur in large errors due to the inherent difficulty and unpredictability of application performance. A system for reliable performance prediction can greatly benefit from the coupling with a local search method based on Petri Nets simulations. This technique allows to obtain very accurate predictions by simulating the whole Hadoop system, clearly at the price of longer execution times, hence making this approach more suitable for design time considerations. In this vision, our models would provide a relevant initial guess for an iterative procedure relying on this more precise technique, in order to find out the optimal configuration or, conversely, certify that an application design will respect the constraints imposed on its execution due to business considerations.